\documentclass[11pt, letterpaper]{article}
\usepackage[utf8]{inputenc}
\usepackage[margin=1.5cm]{geometry}
\usepackage{titlesec}
\usepackage{tabu}
\usepackage{enumitem}
\usepackage{amssymb}
\usepackage{xcolor}
\usepackage{graphicx}
\graphicspath{ {./figs/} }
\newlist{selectlist}{itemize}{2}
\setlist[selectlist]{label=$\square$,leftmargin=*,noitemsep,topsep=0pt}

\usepackage{lmodern}

\usepackage{hyperref}
\hypersetup{
    colorlinks=true,
    linkcolor=blue,
    filecolor=magenta,      
    urlcolor=blue,
}
 
\titleformat{\section}[block]{\hspace{1em}\bfseries}{\thesection.}{0.5em}{} 
\titleformat{\subsection}[block]{\hspace{1em}}{\thesubsection}{0.5em}{}

\begin{document}
\begin{flushleft}

\setlength{\parindent}{0pt}
\setlength{\parskip}{10pt}

\textbf{Modular Light Sources for Microscopy and Beyond (ModLight)}

\textbf{Graham M Gibson$^1$, Robert Archibald$^1$, Mark Main$^2$ and Akhil Kallepalli$^1$*}

\textbf{$^1$ School of Physics and Astronomy, University of Glasgow, Glasgow G12 8QQ United Kingdom}\\ 
\textbf{$^2$ James Watt School of Engineering, University of Glasgow, Glasgow G12 8QQ United Kingdom}

\textbf{Please direct all correspondence to Dr Akhil Kallepalli (Akhil.Kallepalli@glasgow.ac.uk); Twitter: @OptoPhysAkhil}

\textbf{Abstract}\\ Delivering light to an object is one of the key steps in any imaging exercise. Tools such as LEDs and lasers are available to achieve this. These components are integrated into systems such as microscopy, medical imaging, remote sensing, and so many more. Motivated by the need for affordable and open access alternatives that are globally relevant, we share the designs and build instructions for modular light source devices that use simple, off-the-shelf components. Light emitted by near-infrared, red, green and blue LEDs are combined with a choice of mirrors or X-Cube prisms to deliver collimated beams of light. 


\textbf{Specifications table}\\

\vskip 0.2cm

\tabulinesep=1ex
\begin{tabu} to \linewidth {|X|X[3,l]|}
\hline  \textbf{Hardware name} & Modular Light Sources (ModLight)
  \\
  \hline \textbf{Subject area} & %
  \begin{itemize}[noitemsep, topsep=0pt]
  \item Engineering
  \item Physics
  \item Medical (e.g. pharmaceutical science)
  \item Educational tools and open source alternatives to existing infrastructure
  \end{itemize}
  \\
  \hline \textbf{Hardware type} & Imaging tools
  \\ 
\hline \textbf{Closest commercial analog} & No commercial analog is available.
  \\
\hline \textbf{Open source license} &
  CC BY-SA 4.0 (Creative Commons Attribution-ShareAlike 4.0 International).
  \\
\hline \textbf{Cost of hardware} &
  The cost of two devices is approximately £300 (GBP).
  \\
\hline \textbf{Source file repository} & 
    Available with the article.
  \\
\hline \textbf{OSHWA certification UID} \vskip 0.1cm &
In Progress, unavailable as on 07 June 2022.
\\\hline 
\end{tabu}
\end{flushleft}

\section{Hardware in context}
An imaging system's fundamental requirement is a source that produces light, which in turn interacts with the object of interest. This interaction results in phenomena such as refraction, absorption, transmission, scattering and reflection. These interactions, quantified by a detector, result in inferring information regarding the object. The imaging systems are broadly divided into passive (wherein sunlight is used to perform remote sensing from satellite-based imaging systems) and active systems (such as fluorescence microscopy where a light is shone on the object to image it). To effectively image under specific conditions, special light sources and/or control over multiple sources to function is necessary. Such control, currently, is achieved by using different LEDs and laser modules through complex programming or switches that are manually operated. Resources are available to illustrate the importance of choosing a light source\footnote{\url{https://www.scientifica.uk.com/learning-zone/choosing-the-best-light-source-for-your-experiment} \\ \indent \indent \url{https://www.olympus-lifescience.com/en/microscope-resource/primer/lightandcolor/lightsourcesintro/} \\ \indent \indent \url{https://www.rp-photonics.com/white_light_sources.html} \\ \indent \indent \url{https://analyticalscience.wiley.com/do/10.1002/micro.120}} with recent focus on sources that have evolved for specialised imaging with LED sources such as fluorescence microscopy. There is a purpose-built light source for any application, and new sources continue to be engineered to allow better and more accurate imaging. 

New innovations are constantly being presented to the research community, and end users at large. There has been, in recent times, a drive to make research openly accessible and open-source. With the motivation of making light sources more affordable and widely accessible, we introduce two light source designs within the suite of modular light (ModLight) sources that produce collimated light using LEDs. 

The modular light sources use two primary components to direct light, i.e. :
\begin{enumerate}
    \item Mirrors; mounted using magnets and easily switchable to reflect light from the LEDs of choice through the light guide.
    \item X-Cube Prisms; which are oriented to allow light from an LED to transmit through the glass material selectively. Originally designed to combine the three channels into white light, we reverse these processes to transmit light through the light guide. 
\end{enumerate} 

We designed devices that use simple elements to direct near-infrared, red, green and blue light into a light guide, in this instance. The modular nature of the device designs allows any LED to be used for illuminating an object for imaging. Each device can be made with 3D printing and easily accessible parts. This material is intended to be self-sufficient, to the best of our ability, and we are happy to assist where needed. All the latest files, updates and news from the ModLight suite of tools will constantly be updated on \url{https://kallepallilab.com/modular-microscopy/}. 

\section{Hardware description}
The devices made available through this article are specifically designed to facilitate low-cost, robust light sources that can be combined with imaging systems, microscopes, etc. The devices apply the principles of simple optics, along with mirrors and X-Cube prism to direct light for LEDs into a light guide/fibre. The output of these is combined with a lens, resulting in a collimated beam. The devices are 3D printed with custom designs to house the LED assembly along with the mirror/X-Cube prism. The devices are custom-built, with novel designs and benefit from a highly modular nature. For instance, a fluorescence microscopy system is usually fitted with a combination of excitation-emission filters and a light source. Applying the technologies presented in this work, the excitation wavelength can be changed with specific LED modules or the white light source from a combination of light using the X-Cube prism can deliver specific wavelength. Further, these LED modules can be individually built into a suite of sources at specific wavelengths. Should the application need different wavelengths, the LED assemblies can serve as plug-and-play replacements\footnote{All the components from the \textit{Bill of Materials} and 3D printed components from the \textit{Design files summary} are mentioned in \textbf{bold} in the text.}. Concisely, 
\begin{enumerate}
    \item The hardware in this work allow two approaches for delivering collimated light for imaging purposes. 
    \item White light can be realised as a resulting combination of blue, green and red wavelengths using the X-Cube prism. 
    \item The modular nature of the LED assemblies allows for changing illuminating wavelengths rapidly and effectively. 
    \item The custom designs and novel methods of delivering light for an imaging application are 
\end{enumerate}

\section{Design files summary}
\vskip 0.1cm
\tabulinesep=1ex
\begin{tabu} to \linewidth {|X[1.5,1]|X|X[1.5,1]|X[1.5,1]|} 
\hline
\textbf{Design filename} & \textbf{File type} & \textbf{Open source license} & \textbf{Location of the file} \\\hline
\multicolumn{4}{|c|}{\textbf{Common Parts}} \\\hline 
Fibre Collimator & STL CAD file & CC BY-SA 4.0 & Available with the article.  \\\hline
Fibre Holder & STL CAD file & CC BY-SA 4.0 & Available with the article.  \\\hline
Fibre Mount & STL CAD file & CC BY-SA 4.0 & Available with the article.  \\\hline
Heatsink Drill Guide & STL CAD file & CC BY-SA 4.0 & Available with the article.  \\\hline
Heatsink Mount & STL CAD file & CC BY-SA 4.0 & Available with the article.  \\\hline
PMMA Mount & STL CAD file & CC BY-SA 4.0 & Available with the article.  \\\hline
\multicolumn{4}{|c|}{\textbf{Mirror-based Light Source}} \\\hline 
M - Mirror Mount & STL CAD file & CC BY-SA 4.0 & Available with the article.  \\\hline
M - Light Box & STL CAD file & CC BY-SA 4.0 & Available with the article.  \\\hline
M - Light Box (Lid) & STL CAD file & CC BY-SA 4.0 & Available with the article.  \\\hline
M - Source File & OpenSCAD file & CC BY-SA 4.0 & Available with the article.  \\\hline
\multicolumn{4}{|c|}{\textbf{X-Cube prism-based Light Source}} \\\hline 
X - LED Spacer & STL CAD file & CC BY-SA 4.0 & Available with the article.  \\\hline
X - Light Mixer & STL CAD file & CC BY-SA 4.0 & Available with the article.  \\\hline
X - Light Mixer (Lid) & STL CAD file & CC BY-SA 4.0 & Available with the article.  \\\hline
X - Source File & OpenSCAD file & CC BY-SA 4.0 & Available with the article.  \\\hline
\multicolumn{4}{|c|}{\textbf{PCB and LED Control (Electronics)}} \\\hline 
LED Driver (PCB) & PDF & CC BY-SA 4.0 & Available with the article.  \\\hline
LED Driver (Schematic) & PDF & CC BY-SA 4.0 & Available with the article.  \\\hline
LED (Quad Control) & PNG file & CC BY-SA 4.0 & Available with the article. \\\hline
\end{tabu}

\vskip 0.3cm
\noindent
The common components of the both light sources include: 
\begin{enumerate}
    \item Fibre Collimator; the part will house the PMMA lens and the fibre, and deliver the collimated beam as output. 
    \item Fibre Holder; this part holds the fibre against the mount, wherein the PMMA lens is used to collect the light (from the mirror or the X-Cube prism) and focus it into the fibre.
    \item Fibre Mount; the part houses the PMMA lens that collects the light from the mirror/X-Cube prism and attaches to the Fibre Holder. Combined, these three parts make up the fibre assembly. 
    \item Heatsink Drill Guide; The heatsinks (see \textbf{Bill of Materials}\footnote{Available with the article as a separate spreadsheet.}) are solid aluminium components that require drill holes to allow for screws. The positioning of these drill holes is directed by the Heatsink Drill Guide. Only one of these will suffice, and can be reused for all the LEDs assembly. 
    \item Heatsink Mount; this part allows for combining the PMMA Mount with the heatsink to complete the LED assembly.
    \item PMMA Mount; this part holds the PMMA lens and has precisely position screws to hold the LED assembly together as one unit. 
\end{enumerate}

\noindent The mirror-based light source requires the mount (\textbf{M - Mirror Mount}) to hold the mirror (\textbf{Mirror}\footnote{See Bill of Materials, available with the article as a separate spreadsheet.}) in place. The LED assembly and the mirror mount are all places in the box, ``M - Light Box''. The source assembly is completed with a lid (\textbf{M - Light Box (Lid)}). Similarly, the parts for the X-Cube prism-based light source require a compact housing for the LED assemblies and the X-Cube prism placement (\textbf{X - Light Mixer}). Both lids for the source devices are equipment with magnets to allow for a light-tight, completed build. All components are available and editable through the OpenSCAD files (M - and X - Source File). Finally, the LEDs are controlled using a custom PCB board requiring drivers (\textbf{LED-IC}) and potentiometers (\textbf{LED-P}); the details of the PCB board and the schematic for the driver are shared in the PDF files, \textbf{LED Driver (PCB)} and \textbf{LED Driver (Schematic)}. 

\begin{figure}[t]
\centering
\includegraphics[width=0.95\columnwidth]{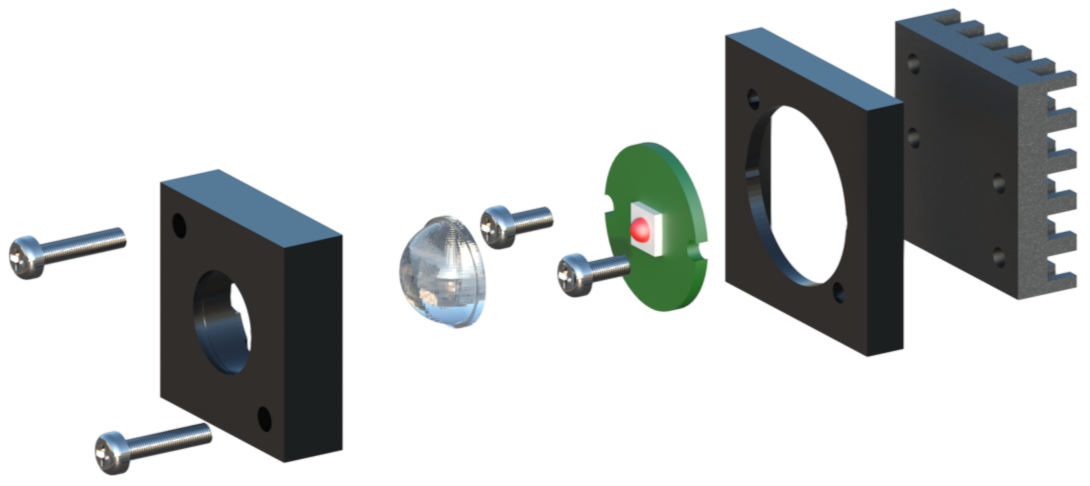}
\caption{The LED module, common to both sources, can be assembled using 4 screws (2x \textbf{S-M2.5-12}, 2x \textbf{S-M2.5-6}), the lens mount (\textbf{PMMA-f5}), LED modules (\textbf{LED-IR-xxx}, \textbf{LED-R}, \textbf{LED-G}, \textbf{LED-B}), a heatsink mount (\textbf{Heatsink Mount}) and a heatsink (\textbf{HS}). A guide (\textbf{Heatsink Drill Guide}) is available for accurate screw hole placements in the heatsink.}
\label{fig:LED_Module}
\end{figure} 

\begin{figure}[t]
\centering
\includegraphics[height=6cm]{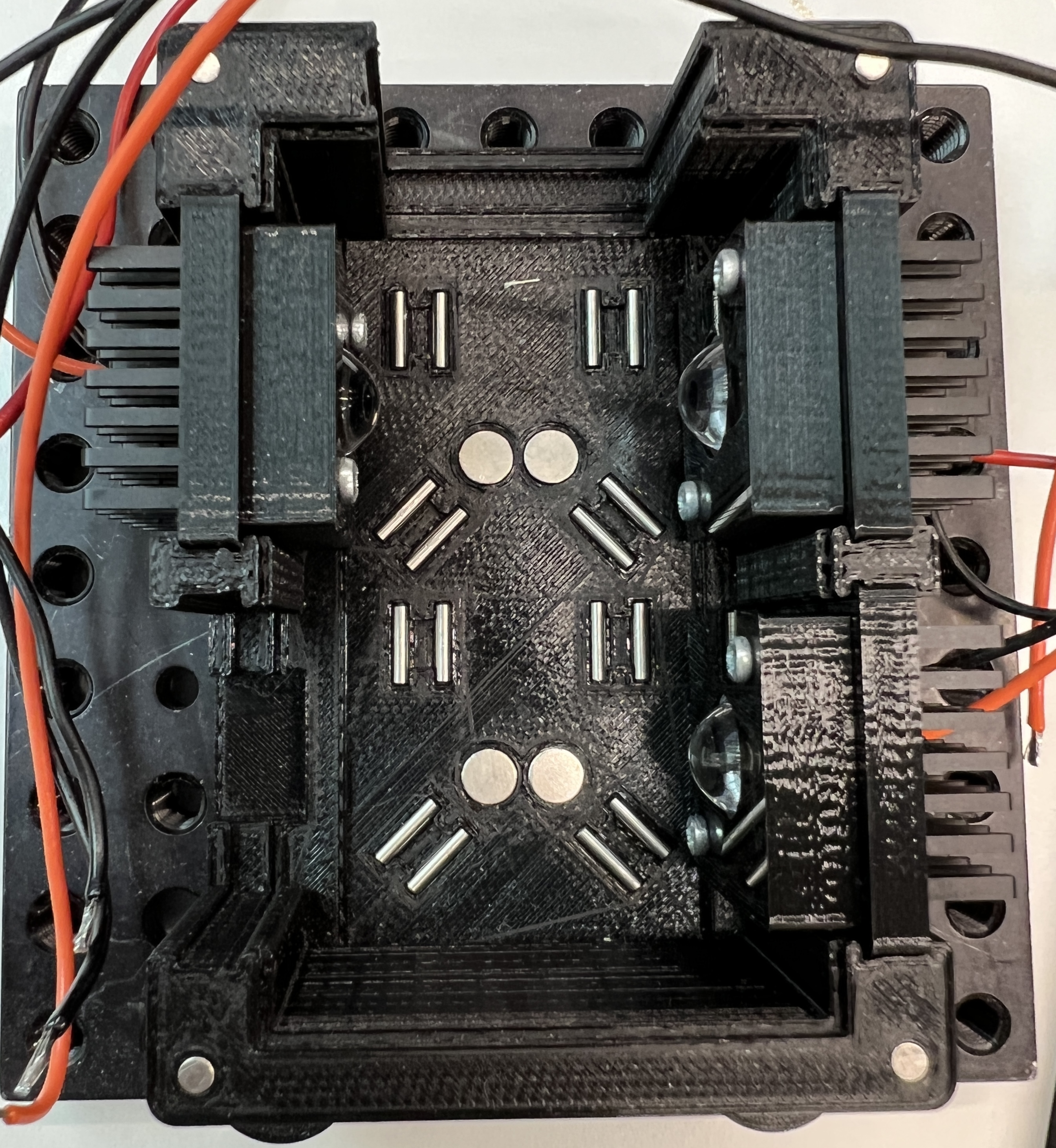}
\includegraphics[height=6cm]{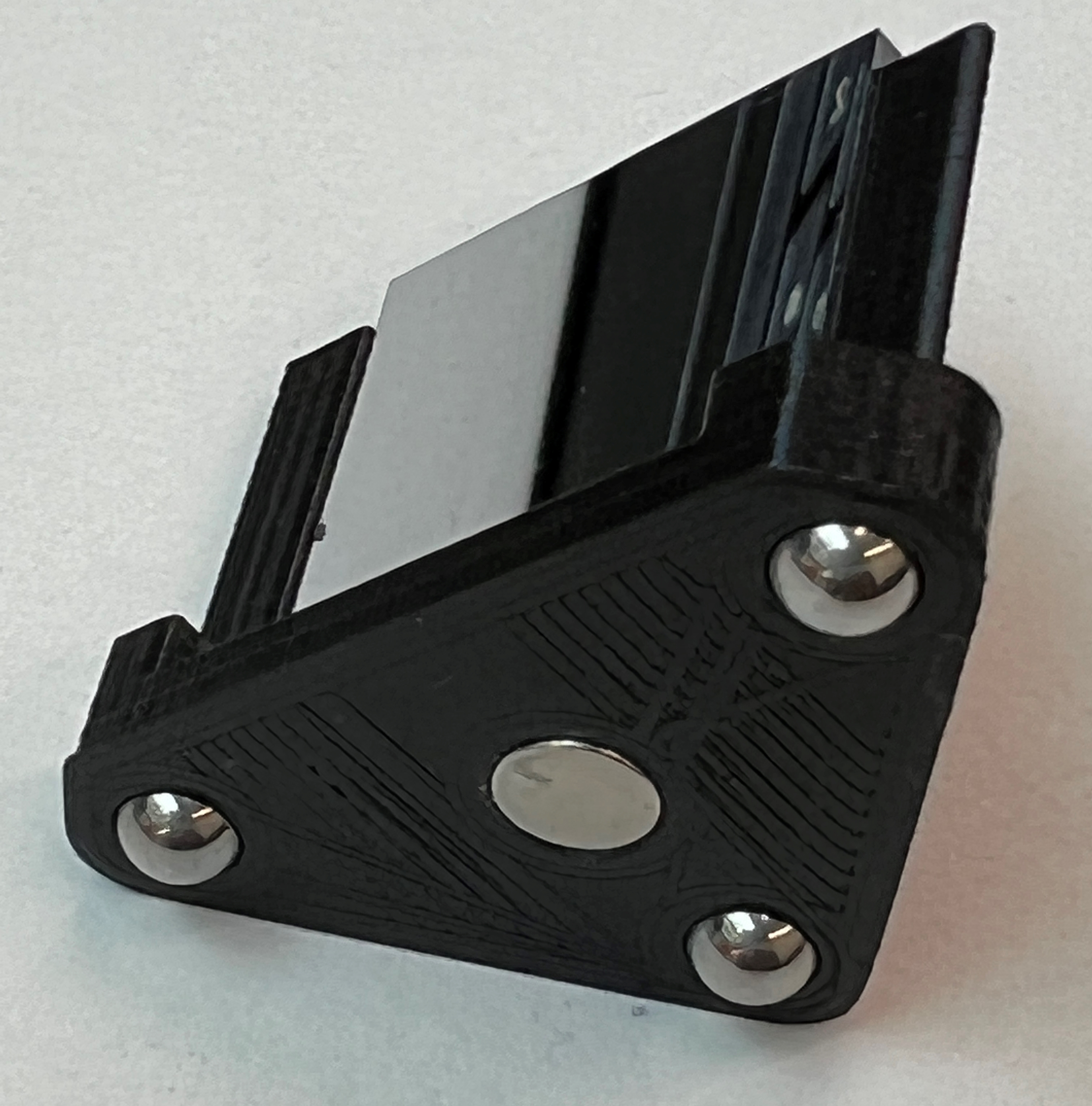} 
\includegraphics[height=6cm]{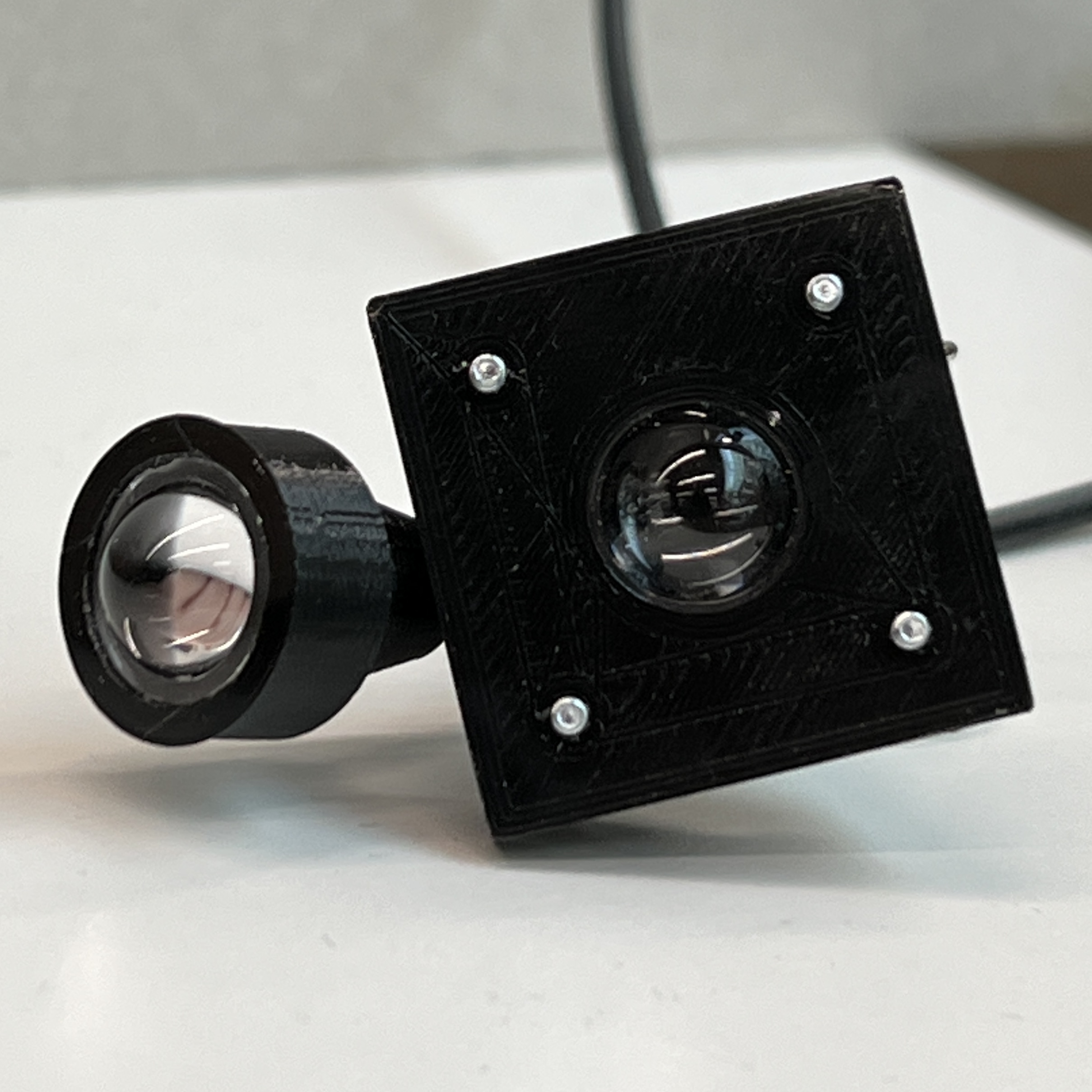}
\caption{Before assembling all the components of the light source, the base of the light box (\textit{left}), the mirror mount (\textit{middle}) and collimated light delivery through the light guide (\textit{right}) should be prepared. The dowel pins (\textbf{P-2x8}) and magnets (\textbf{M-6x3}) for the light box; and ball bearings (\textbf{BB-5 mm}) and magnets (\textbf{M-6x3}) for the mirror mount are illustrated, assembled and ready for use. The combination of dowel pins-ball bearings, and magnets ensure that the mirror is aligned at 45\textdegree \ to the LED of choice.}
\label{fig:Base_Mirror_Fibre_Prep}
\end{figure} 

\section{Bill of materials}
For a detailed bill of materials, please find available an editable spreadsheet uploaded with this manuscript. 

\section{Build instructions}
The 3D printing of all the components takes a few days. We recommend beginning the 3D printing while ordering and awaiting delivery of the various components from the Bill of Materials. The build process for both sources have three common sub-builds:
\begin{enumerate}
    \item LED assembly (Section \ref{subsec:LED})
    \item Fibre-coupled collimated beam delivery (Section \ref{subsec:Fibre})
    \item PCB and Electronics for LED control (Section \ref{subsec:PCB})
\end{enumerate}

\noindent After the common components have been built, either the mirror mount-based (Section \ref{subsec:Mirrors}) or the X-Cube prism-based device (Section \ref{subsec:XCube}) can be assembled. 

\subsection{\textbf{LED Assembly}} \label{subsec:LED}
The LED assembly is modular, and can be completed with any LED at any wavelength, as long as the LED module fits within the set up. We envisage that specific applications can be supported by building a suite of LED assemblies, each of which can be replaced/changed in the source device. 

The LED assembly requires 3D printed parts (\textbf{Heatsink Drill Guide}, \textbf{Heatsink Mount}, \textbf{PMMA mount}) and purchased components (\textbf{HS}, \textbf{LED-IR-xxx}, \textbf{LED-x}, \textbf{PMMA-f5}, \textbf{S-M2.5-6}, \textbf{S-M2.5-12}). The process can be visualised from the illustration, Figure \ref{fig:LED_Module}. The LED assembly can be completed through the following steps:
\begin{enumerate}
    \item Collect the 3D printed parts. Place the \textbf{Heatsink Drill Guide} on top of the heatsink (\textbf{HS}) to make screw holes in the precise locations. 
    \item Once complete, mount the \textbf{LED-x} (choice of LED) on top of the heatsink and wire it appropriately using the red and black cables (\textbf{EW-R}, \textbf{EW-B}). Allow for a sufficient length of cables, depending on your workspace and application. 
    \item Use the 6 mm M2.5 screws (\textbf{S-M2.5-6}) to secure the LED module to the heatsink. Place the \textbf{Heatsink Mount} on top to hold the LED and wires in place. 
    \item Take a 5 mm focal length PMMA lens (\textbf{PMMA-f5}) and place it securely inside the \textbf{PMMA Mount}. Once completed, align the screw holes and complete the assembly by securing all the components with the 12 mm M2.5 (\textbf{S-M2.5-12}) screws.
\end{enumerate}

\subsection{\textbf{Fibre-coupled collimated beam delivery}} \label{subsec:Fibre}
The light that has been directed either by the plane mirror or the X-Cube prism towards the output requires collection and flexible delivery. Further, the output must be a collimated beam for applications such as fluorescence microscopy. Therefore, as described below in the previous section, the common components can be combined to collect and deliver the light to the imaging application/experiment. 
\begin{enumerate}
    \item Collect the 3D printed parts (\textbf{Fibre Mount}, \textbf{Fibre Holder}, \textbf{Fibre Collimator}) and the purchased components (\textbf{PMMA-f5}, \textbf{S-M2.5-12}, \textbf{Fibre} of a suitable length). 
    \item Place the 5 mm focal length lens in the \textbf{PMMA Mount}. This part of the set up collects the light from the mirror or the X-Cube prism. The mount is then aligned with the \textbf{Fibre Holder} which holds one end of the light guide (\textbf{Fibre}). 
    \item The other end of the light guide fits into the \textbf{Fibre Collimator}. Before fitting the light guide into the collimator, place the PMMA lens (\textbf{PMMA-f5}) on the wider end. Combined with the light guide and the lens, a collimated beam output is realised (Figure \ref{fig:Base_Mirror_Fibre_Prep} (\textit{right})). 
\end{enumerate}

\begin{figure}[h]
\centering
\includegraphics[height=7.5cm]{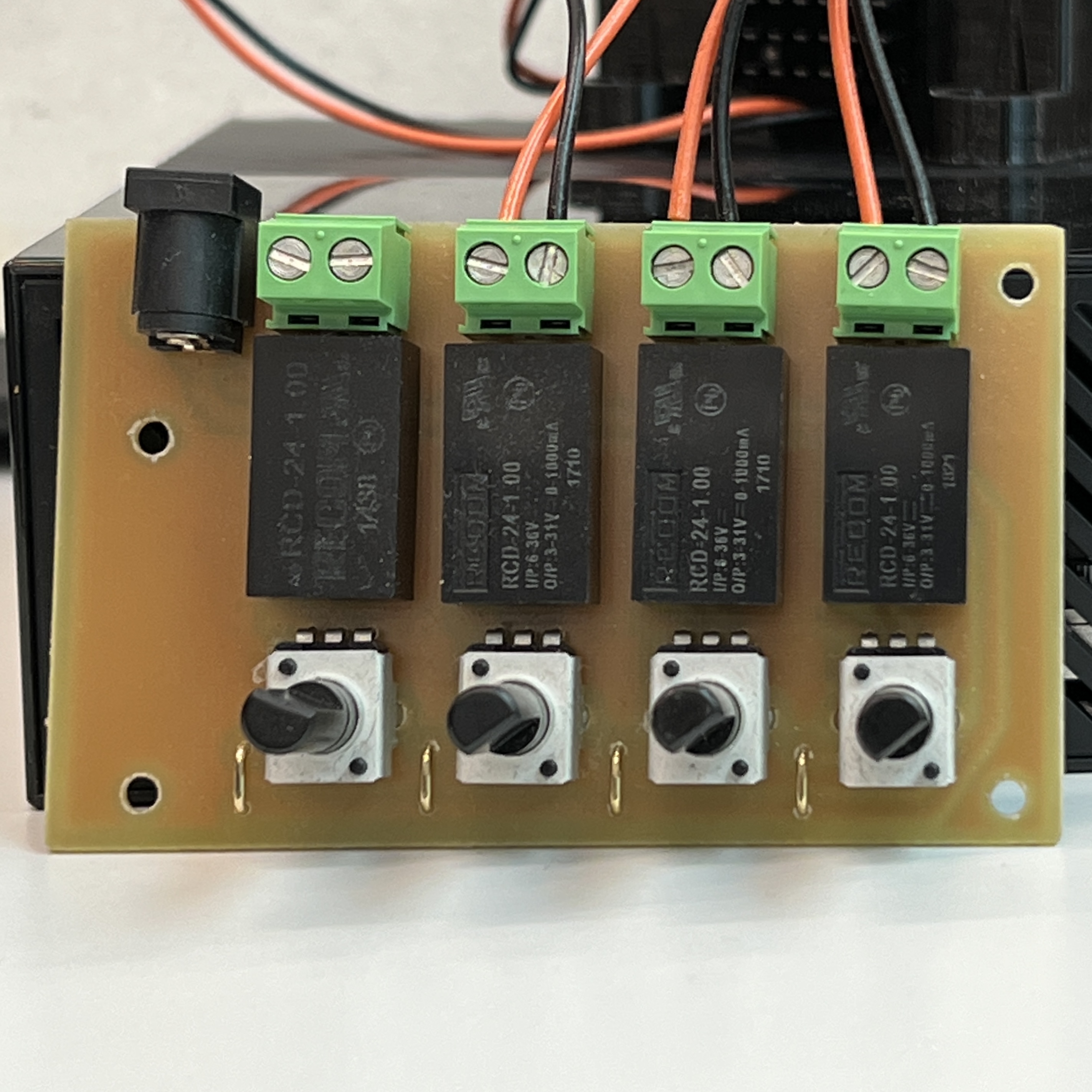} \\
\includegraphics[width=0.8\columnwidth]{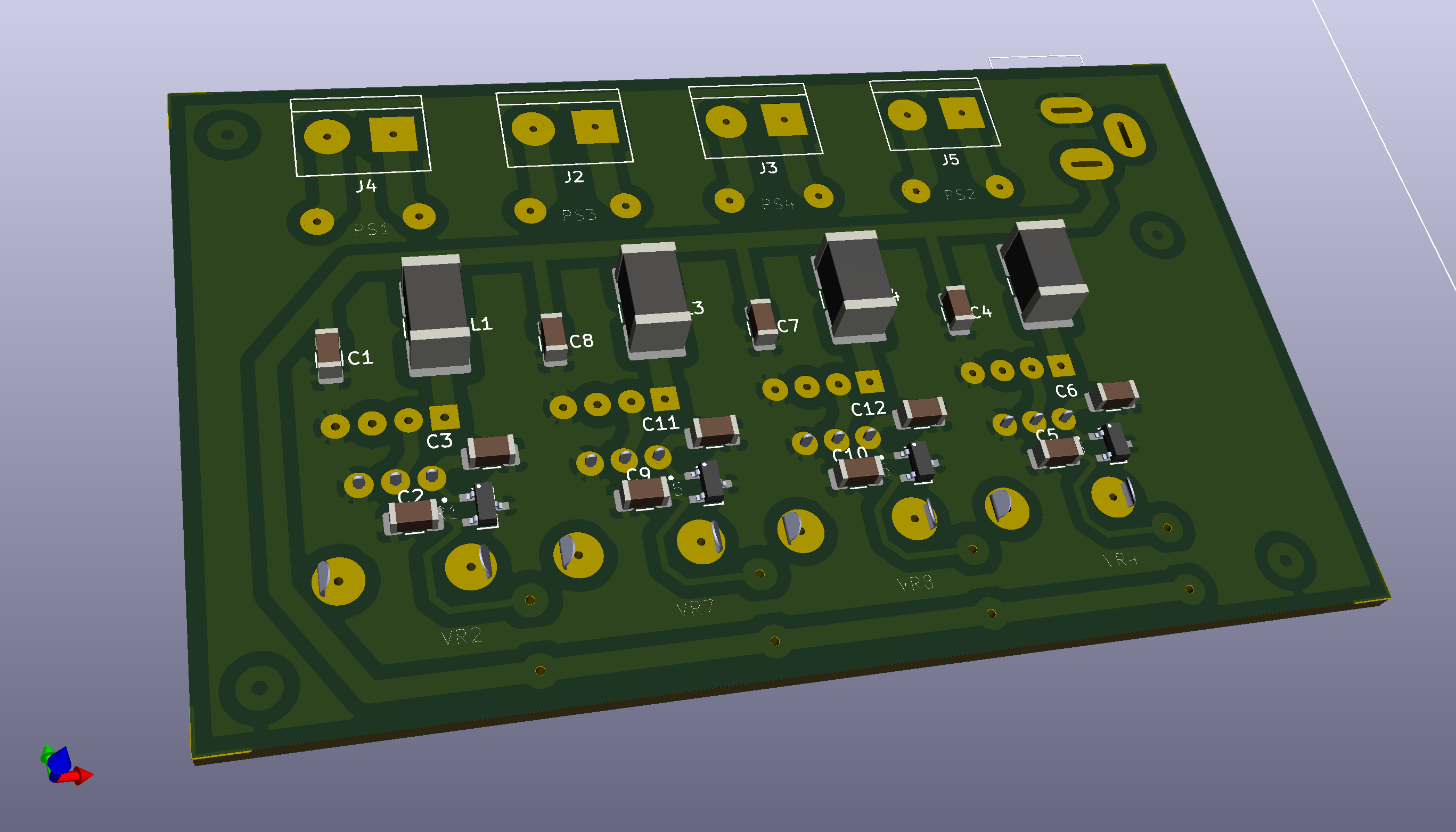} 
\caption{The LEDs are controlled using 4 potentiometers (\textbf{LED-P}) and drivers (\textbf{LED-IC}) as illustrated (\textit{top}), with the board design also shown (\textit{bottom})}
\label{fig:LED_Electronics}
\end{figure} 

\subsection{\textbf{PCB and Electronics}} \label{subsec:PCB}
The electronic circuit schematic for the LED controller is shared in the PDF file \textbf{LED Driver (Schematic)}. The circuit is based on the RCD-24-1.00 LED controller module with only a small number of additional components. A PCB board design is shared in the PDF file \textbf{LED Driver (PCB)}. The drivers (\textbf{LED-IC}) and potentiometers (\textbf{LED-P}) are placed on the board, and the LEDs are wired in to the controls. This design offers control beyond simply turning the LEDs on and off. The control board and design are illustrated in figure \ref{fig:LED_Electronics}.

\begin{figure}[t]
\centering
\includegraphics[width=0.95\columnwidth]{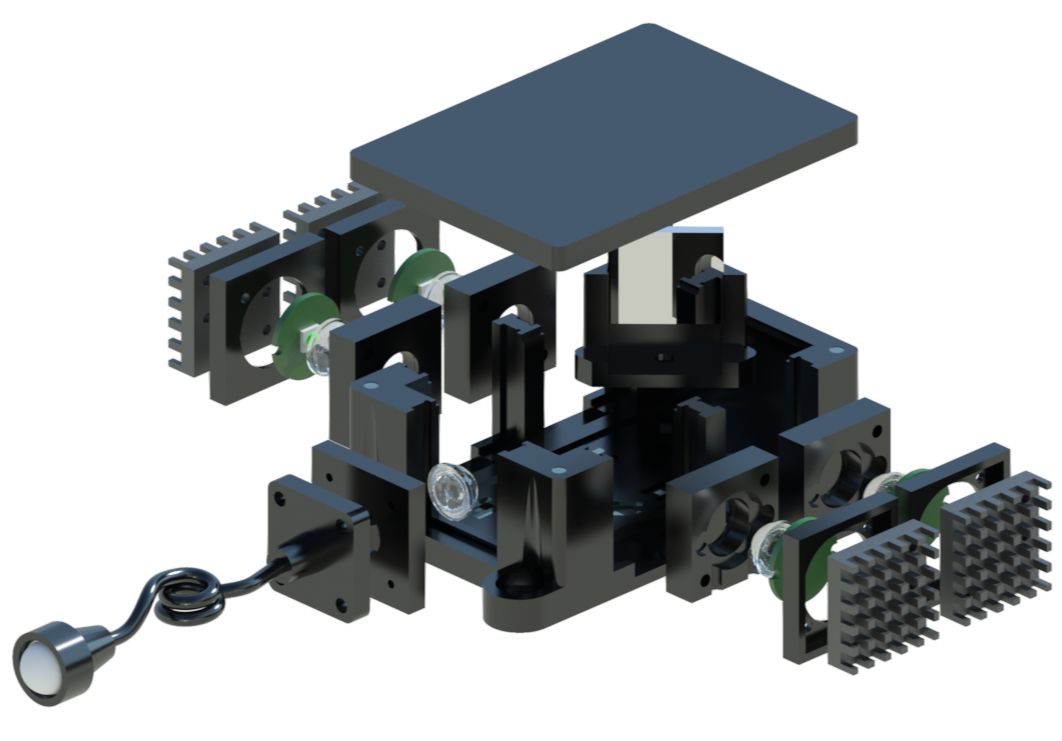}
\caption{The LED modules can be integrated into the main device using other 3D printed parts. The components and parts needed are the light box (\textbf{M - Light Box}) and lid (\textbf{M - Light Box (Lid)}), magnets (\textbf{M-6x3}, \textbf{M-3x2}), dowel pins (\textbf{P-2x8}), mirror mount (\textbf{M - Mirror Mount} and a mirror (\textbf{Mirror}) The ball bearings underneath the mirror mount allow easy place. Once assembled, the light delivery component can be assembled with the combination of a lenses (2x \textbf{PMMA-f5}), fibre mount (\textbf{Fibre Mount}), fibre holder (\textbf{Fibre Holder}) and fibre collimator (\textbf{Fibre Collimator}). The LED modules are wired to a simple switch that can turn each LED on and off as required.}
\label{fig:MMount_ExplodedView}
\end{figure} 

\begin{figure}[]
\centering
\includegraphics[width=0.95\columnwidth]{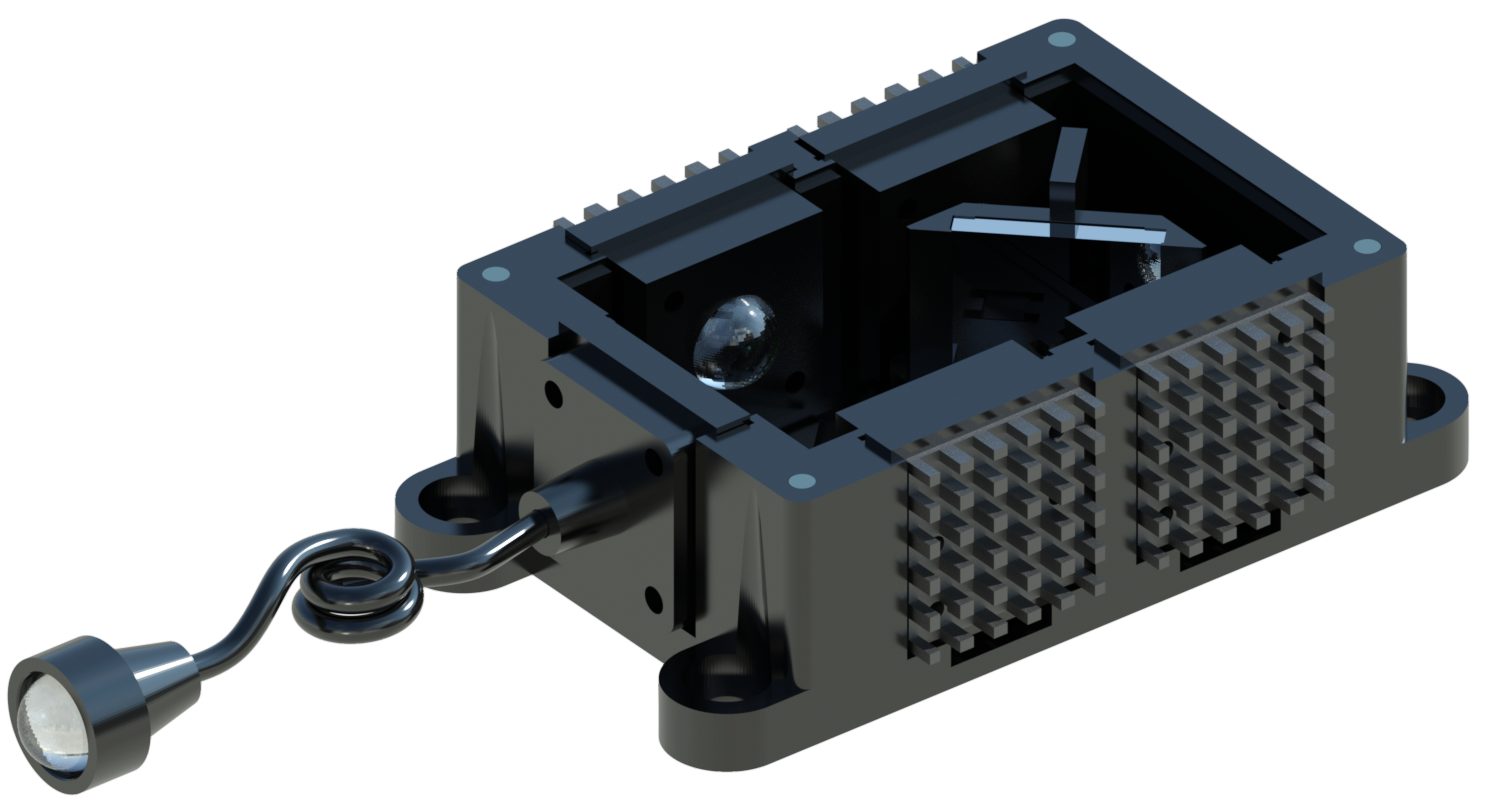}
\caption{The mirror-mount source allows individual control of LEDs, directing the light from the source to the fibre and through the collimator. The lid of the device can be lifted to move the magnet-supported mount can be into any of the 4 positions corresponding to the LEDs at 45\textdegree \ angles. Once the mirror is in the position of choice, the lid can be replaced for a light-tight source with a collimated beam as output from the collimator.}
\label{fig:MMount_UnexplodedView}
\end{figure} 

\subsection{\textbf{Devices with Mirrors}} \label{subsec:Mirrors}
The LED assemblies can be placed in the slots on the vertical walls of the main housing, \textbf{M - Light Box}. The detailed illustration of the device is given in the exploded (Figure \ref{fig:MMount_ExplodedView}) and completed views (Figure \ref{fig:MMount_UnexplodedView}). The light box can be assembled as follows: 
\begin{enumerate}
    \item Once printed, place the dowel pins (\textbf{P-2x8}) in their respective slots on the base of the box. 
    \item Place and secure the magnets (\textbf{M-6x3}) on the base. Ensure that the correct poles of the magnets are aligned by pairing magnets before securing them in place. The four magnets in the base of the box need to be of the opposite polarity as that of the mirror mount. 
    \item Take the mirror mount (\textbf{M - Mirror Mount}) and secure three ball bearings (\textbf{BB-5mm}) and a magnet (\textbf{M-6x3}) at its base. 
    \item Place the LED modules in the slots and the mirror (\textbf{Mirror}) in the mirror mount in one of the positions to secure it in place. 
    \item Finally, place four \textbf{M-3x2} magnets in top corners of the light box (\textbf{M - Light Box}) and four magnets of opposite polarity in the lid, \textbf{M - Light Box (Lid)}. 
    \item The fibre coupling assembly described in the previous subsection, \textit{Fibre-couple collimated beam delivery}, can now be placed in the appropriate slot. 
    \item Place the lid over the box, secured by magnets to complete a light-tight, 4-LED source that can deliver collimated light through the output end of the light guide. 
\end{enumerate}

\subsection{\textbf{Devices with X-Cubes}} \label{subsec:XCube}
\begin{figure}[t]
\centering
\includegraphics[width=0.95\columnwidth]{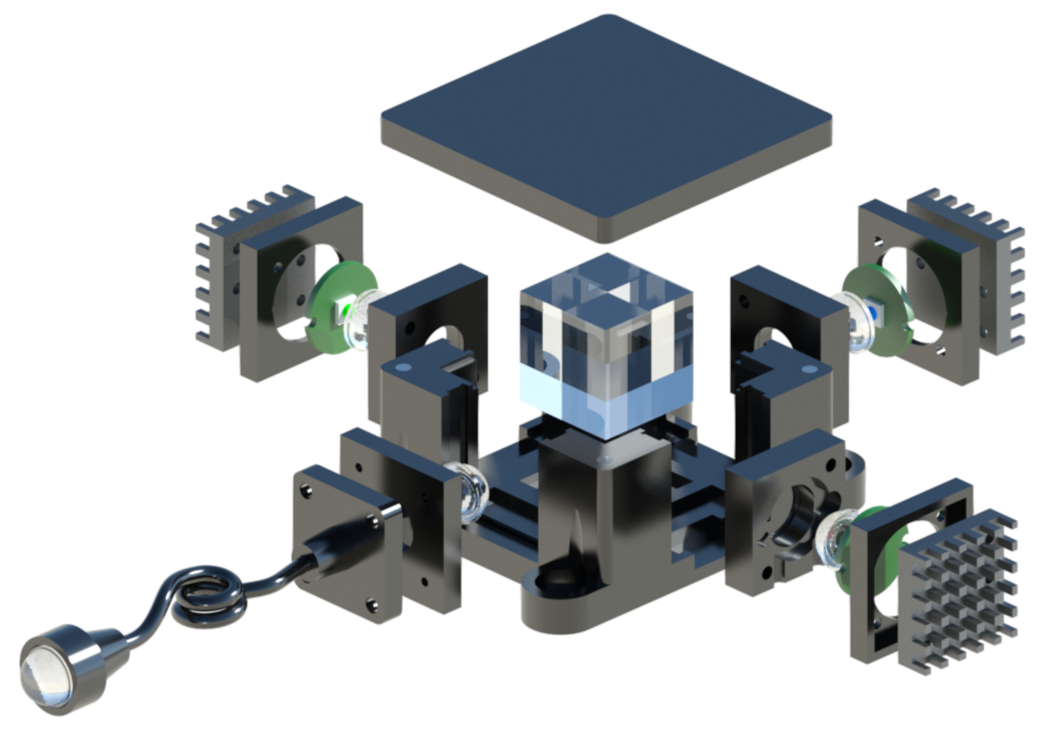}
\caption{The LED modules can be integrated into the main device using other 3D printed parts. The components and parts needed are the light box (\textbf{X - Light Mixer}) and lid (\textbf{X - Light Mixer (Lid)}), magnets (\textbf{M-3x2}) and an X-Cube (\textbf{XCube}). Once assembled, the light delivery component can be assembled with the combination of a lenses (2x \textbf{PMMA-f5}), fibre mount (\textbf{Fibre Mount}), fibre holder (\textbf{Fibre Holder}) and fibre collimator (\textbf{Fibre Collimator}). The LED modules are wired to a simple switch that can turn each LED on and off as required}
\label{fig:XCube_ExplodedView}
\end{figure}

\begin{figure}[t]
\centering
\includegraphics[width=0.95\columnwidth]{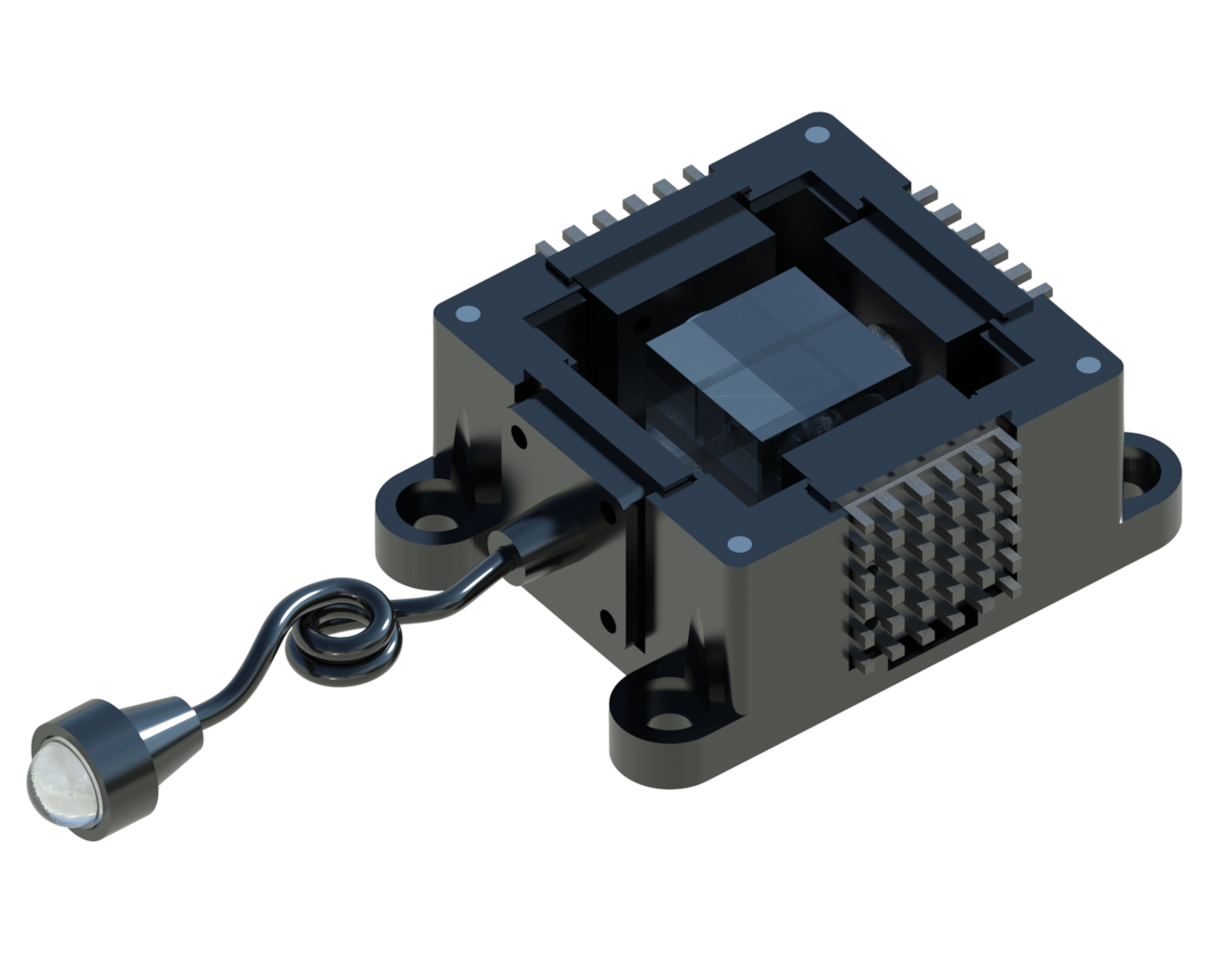}
\caption{The X-Cube prism-based source allows individual control of LEDs, directing the light from the source to the fibre and through the collimator. This design is significantly more compact than the mirror-based source, and allows for compatibility with fluorescence microscopy systems that integrate multiple light sources and filter cube combinations, without needing to change sources. This single source can combine multiple broadband sources simultaneously, and is capable of delivering appropriate excitation wavelengths to the sample. Once the X-Cube prism is in place, the lid can be replaced for a light-tight source with a collimated beam as output from the collimator.}
\label{fig:XCube_UnexplodedView}
\end{figure}

The design of the X-Cube prism-based source allows for a much more compact design and simultaneous operation of multiple wavelengths. Either individual LEDs can be used for a ``pure'' output or multiple wavelengths can be combined into a single source and the output can be filtered in the experiment; the choice is available to the user. The detailed illustration of the device is given in the exploded (Figure \ref{fig:XCube_ExplodedView}) and completed views (Figure \ref{fig:XCube_UnexplodedView}). The device can be assembled in the following steps: 
\begin{enumerate}
    \item Once printed, secure the X-Cube prism (\textbf{Xcube}) in the correct position, aligning the LEDs appropriately to the correct sides of the prism. 
    \item Place the LED modules in the slots. 
    \item Finally, place four \textbf{M-3x2} magnets in top corners of the light box (\textbf{X - Light Mixer}) and four magnets of opposite polarity in the lid, \textbf{X - Light Mixer (Lid)}. 
    \item The fibre coupling assembly described in the previous subsection, \textit{Fibre-couple collimated beam delivery}, can now be placed in the appropriate slot. 
    \item Place the lid over the box, secured by magnets to complete a light-tight, 3-LED source that can deliver collimated light through the output end of the light guide. 
\end{enumerate}

\noindent
\textbf{CRediT author statement}\\

\noindent
\textbf{Graham Gibson}: Conceptualisation, Methodology, Validation, Resources, Data Curation, Writing - Review \& Editing \textbf{Robert Archibald}: Methodology, Validation \textbf{Mark Main}: Visualisation, Writing - Review \& Editing \textbf{Akhil Kallepalli}: Methodology, Validation, Investigation, Writing - Original Draft, Writing - Reviewing \& Editing \\

\noindent
\textbf{Acknowledgements}\\

\noindent This work was made possible with support from the Research Fund awarded by the Endowments Sub Committee of the British Association of Oral and Maxillofacial Surgeons (2021), Engineering and Physical Sciences Research Council (EPSRC) Impact Acceleration Account (IAA) [EP/R511705/1]; EPSRC Funding to QuantIC [EP/M01326X/1], University of Glasgow [Glasgow Knowledge Exchange Fund]; and the Royal Society. \\


\end{document}